\documentclass[12pt]{article}
\def\cF{{\cal F}}

\newfont{\goth}{eufm10 scaled \magstep1}

\def\a{\alpha}

\def\b{\beta}

\def\c{\gamma}
\def\d{\delta}
\def\e{\epsilon}

\def\l{\lambda}

\def\r{\rho}
\def\s{\sigma}
\def\t{\tau}

\def\beq{\begin{equation}}\def\eeq{\end{equation}}
\def\beqa{\begin{eqnarray}}\def\eeqa{\end{eqnarray}}
\def\barr{\begin{array}}\def\earr{\end{array}}

\def\del{\partial}\def\delb{\bar\partial}

\def\dt{\tilde{\d}}

\def \ys {{y\kern-.5em / \kern.3em}}

\let\bm=\bibitem

\def\nn{\nonumber}
\def\bd{\begin{document}}
\def\ed{\end{document}}
\def\ba{\begin{array}}
\def\ea{\end{array}}
\def\bea{\begin{eqnarray}}
\def\eea{\end{eqnarray}}
\def\ft#1#2{{\textstyle{{\scriptstyle #1}\over {\scriptstyle #2}}}}
\def\fft#1#2{{#1 \over #2}}
\newcommand{\be}{\begin{equation}}
\newcommand{\ee}{\end{equation}}
\newcommand{\eq}[1]{(\ref{#1})}
\def\eqs#1#2{(\ref{#1}-\ref{#2})}
\def\det{{\rm det\,}}
\def\tr{{\rm tr}}
\newcommand{\ho}[1]{$\, ^{#1}$}
\newcommand{\hoch}[1]{$\, ^{#1}$}
\def\ra{\rightarrow}
\def\uha{{\hat {\underline{\a}} }}
\def\uhc{{\hat {\underline{\c}} }}
\def\n{{\rm n}}
\def\m{{\rm m}}
\def\k{{\rm k}}

\def \Om {\Omega}
\def \bfd {{\bf d}}
\def \del {\partial}
\def \eps {\epsilon}
\def \Z {{\bf Z}}
\def \xb {\bar{x}}
\def \la {\langle}
\def \ra {\rangle}
\def \Omt {\tilde \Omega} 
\def \la {\langle}
\def \ra {\rangle}
\def \psib{\overline{\psi}}

\def \II {I\hspace{-.1em}I\hspace{.2em}}
\begin{document}

\hfill{NEIP-99-011}

\hfill{hep-th/9906192}

\vspace{20pt}

\begin{center}

{\Large\bf Constrained Quantization of  Open String 
in Background $B$ Field  and 
Noncommutative D-brane}
\vspace{30pt}

{\large Chong-Sun Chu\hoch{1} and Pei-Ming Ho\hoch{2}}

\vspace{15pt}

\begin{itemize}
\item[$^1$] {\small \em Institute of Physics, University of
Neuch\^atel, CH-2000 Neuch\^atel, Switzerland}
\item[$^2$] {\small \em Department of Physics, National Taiwan
University, Taipei 10764, Taiwan, R.O.C.}
\end{itemize}

\vskip .2in \sffamily{cschu@sissa.it \\ pmho@phys.ntu.edu.tw}

\vspace{60pt}

{\bf Abstract}
\end{center}

In a previous paper we provided a consistent quantization of open
strings ending on D-branes with a background $B$ field. In this
letter, we show that the same result can also be obtained using the 
more traditional method of Dirac's constrained quantization. 
We also extend the discussion to the fermionic sector.

\newpage

\section{Introduction}

It was shown by Connes, Douglas and Schwarz \cite{CDS}
that for the Matrix model compactified on a torus,
the three-form $C$ field background should be incorporated
in the supersymmetric Yang-Mills theory (SYM) by deforming
the base space into a quantum space.
Due to the correspondence between the DLCQ M theory and
the D-brane worldvolume field theory \cite{Seiberg0},
this means that the D-brane field theory lives on
a noncommutative space in the presence of
a NS-NS $B$ field background \cite{DH,HV,Seiberg}.
In a previous paper we quantized an open string ending
on a D-brane with a constant NS-NS $B$ field background,
and we found that the spacetime coordinates of the open
string endpoint are noncommutative.
The noncommutativity precisely agrees with
previous results \cite{CDS}, but in addition we showed
that the 
%c1
compactness of the torus 
%c1
is not necessary for
the noncommutitivity, and in general the deformation
is parametrized by $\cF=B-F$ instead of $B$.

The quantization of an open string we did in \cite{CH}
was carried out explicitly in terms of a mode expansion of
the generic solution of the equations of motion
for the spacetime coordinates $X^{\mu}$.
We used a heuristic argument in the derivation of
commutation relations among the mode coefficients,
and then check that it is
a consistent quantization of the classical theory.
%Recently it is pointed out by Seiberg \cite{Seiberg}
%that there are more than one possible quantizations
%of the same system, corresponding to different
%regularizations of the theory.
%In particular the quantization with a noncommutative
%base space is better in the DLCQ limit in the sense
%that the SYM theory does not have to be corrected by
%higher order terms.

In this paper we carry out the quantization again
by following the standard quantization procedure
of Dirac for systems with constraints.
The constraints here are, of course, the boundary conditions
of the open string ending on a D-brane.
We find that there is a single ambiguous expression which needs
to be regularized.
% p2 in the quantization procedure.
We obtain our previous results by choosing a natural regularization.
We also find the commutation relations for the fermionic fields
using supersymmetry.

\section{Necessity of Noncommutativity} \label{necess}

We first review briefly the setting.  Consider a fundamental string
ending on a D$p$-brane.
This can be in type \II superstring, type 0 superstring, or in the bosonic
string theory. The bosonic part of the action takes the same form
\cite{DLP,Lei} 
\be \label{action} S_B= {1 \over 4\pi\alpha'}
\int_{\Sigma} d^2\sigma \bigl[ g^{\a\b}G_{\mu\nu} \partial_\a
X^{\mu}\partial_\b X^{\nu}+ \eps^{\a\b} B_{\mu\nu}\partial_\a
X^{\mu}\partial_\b X^{\nu} ] + {1 \over 2\pi\alpha'}\oint_{\partial
\Sigma} d \tau A_i(X) \partial_{\tau}X^i, 
\ee 
where $A_i,\ i=0,1,\cdots, p$, is the $U(1)$ gauge 
field living on the D$p$-brane.
We use the convention $\eta^{\a\b}=\mbox{diag}(-1,1)$ and
$\eps^{01}=1$ as in \cite{CH}.  Here the string background is 
\be \label{flat}
G_{\mu\nu} = \eta_{\mu\nu}, \quad \Phi =\mbox{constant}, \quad H=dB=0.
\ee 
Adding the fermionic sector does not affect the result
% p2 in this section
and will be considered later.
% p2 in the next section. 
With slight modification, the considerations in this paper 
can also be applied to study open string ending on a D-brane in
type I string theory.

If both ends of a string are attached to the same D$p$-brane, the last
term in (\ref{action}) can be written as 
\be
\frac{-1}{4\pi\a'}\int_{\Sigma}d^2\s
\eps^{\a\b}F_{ij}\del_{\a}X^i\del_{\b}X^j.  
\ee 
Furthermore, consider
the case $B=\sum_{i,j=0}^{p}B_{ij}dX^i dX^j$, then the action
(\ref{action}) can be written as 
\be 
S_B=-\int d\t
L=\frac{1}{4\pi\a'}\int d^2\s \bigl[ g^{\a\b}\eta_{\mu\nu}\partial_\a
X^{\mu}\partial_\b X^{\nu}+ \eps^{\a\b}\cF_{ij}\partial_\a
X^{i}\partial_\b X^{j} ].  
\ee
% p1
Here 
\be \cF =B-dA=B-F 
\ee 
is the modified Born-Infeld field strength and $x_0^a$
is the location of the D-brane.  Indices are raised and lowered by
$\eta_{ij} = (-, +, \cdots, +)$.
 
One obtains the equations of motion 
\be 
(\del^2_{\tau}-\del^2_{\s})
X^\mu =0 \ee and the boundary conditions at $\s =0, \pi$: \bea
&\del_\s X^i + \del_\tau X^j \cF_j{}^i =0, \quad i,j= 0,1,\cdots, p,
\label{BC1}\\
&X^a =x_0^a,\quad a =p+1, \cdots, D. \label{BC2} 
\eea
 
The constraint \eq{BC2} is standard.  We will be mainly 
interested in the
constraint \eq{BC1}.  As demonstrated in \cite{CH}, the BC \eq{BC1}
implies that 
\be \label{dXwithP} 
2\pi\a' P^k(\t,0)\cF_k{}^i =-\del_{\s}X^j(\t,0)M_j{}^i, 
\ee 
where $P^k$ is the canonical momentum 
\be \label{mom}
2\pi\a' P^k(\t,\s)=\del_\tau X^k + \del_\sigma X^j\cF_j{}^k,
\ee 
and $M_{ij}=\eta_{ij}-\cF_i{}^k\cF_{kj}$.  
It follows that
\be\label{PP1}
2\pi\a'[P^k(\t,0),P^j(\t,\s')]\cF_k{}^i= 
-\del_{\s} [X^k(\t,\s),P^j(\t,\s')]_{\s=0} \, M_k{}^i, 
\ee 
and 
\be \label{XX1}
2\pi\a'[P^k(\t,0),X^j(\t,\s')] \cF_k{}^i = 
- \del_\s [X^i(\t,\s), X^j(\t,\s')]_{\s=0}.  
\ee

These simple relations show that the standard canonical commutation
relations for $\cF=0$,
\bea 
&[X^i(\t,\s),P_j(\t,\s')]=i\d^i_j\d(\s,\s'), \label{stdCR1} \\
&[P_i(\t,\s),P_j(\t,\s')]=0,\label{stdCR2} \\ 
&[X^i(\t,\s),X^j(\t,\s')]=0,  \label{stdCR3}
\eea 
are no longer valid when $\cF \neq 0$. They are not 
compatible with the boundary condition \eq{BC1} when 
$\cF\neq 0$. In particular, without doing any further calculations, 
one can see already from \eq{XX1} that the string
coordinates must necessarily be noncommutative somewhere along the
string. A
consistent quantization was therefore proposed in \cite{CH} and it was
shown that the canonical commutation relations are modified to 
\bea
&[P^i(\t,\s),P^j(\t,\s')] =0, \label{F1} \\
&[X^k(\t,\s),X^l(\t,\s')] = \left\{
\begin{array}{ll}
\pm 2\pi i \a'( M^{-1} \cF )^{kl}, & \s=\s' =0 \mbox{ or } \pi,\cr 
0, &  \mbox{otherwise},  
\end{array}
\right. \label{F2} \\
&[X^i(\t,\s),P^j(\t,\s')]= i \eta^{ij} \dt(\s,\s'), \label{F3}
\eea
where
$\dt(\s,\s')$ is the delta function on $[0,\pi]$ with vanishing
derivative at the boundary, e.g. $\del_{\s} \dt (\s, \s') =0$ for 
$\s =0, \pi$. See \cite{CH} for its explicit form.
Thus we see that the string becomes noncommutative at the endpoint, 
i.e. the D-brane becomes noncommutative. 
It was further shown that for the case $\cF =B$, 
\eq{F2} agrees with the results obtained from other
considerations \cite{others}.
% p1: They were exact.
%in the leading order of $B$.
The % p2 nonlinear contributions in
result of our quantization was later confirmed
in \cite{Schom} using standard string perturbation theory. 

Although our quantization in \cite{CH} is entirely consistent and
produces results in agreement with other considerations, the argument
there was mainly based on intuitive guess and consistency argument. 
It may perhaps be more satisfactory to give 
a ``derivation'' of that result based on more traditional method. 
In the next section, we will show
that the result in \cite{CH}
% p1
% is entirely equivalent to
can also be obtained from
the standard constraint quantization of Dirac.
Consequences of the quantization \eq{F1}-\eq{F3} 
in the boundary state formalism 
%c1 
(closed string) 
%c1 
and D-brane physics are under investigation \cite{work}; 
%c1
as well as the quantization of charged open string \cite{charged}
and its possible consequences
%c1
\footnote{See in particular the  papers for some
recent related developments:   
\cite{a} studies the closed string interaction in the presence of
noncommutative D-branes;  the papers \cite{b} study the effects of
$B$-field  in the  boundary state formalism.  }.

\section{Dirac Quantization of the Bosonic Sector}

In this section, we compute the Dirac bracket 
starting from the standard Poisson brackets
\bea
&(X^i(\s),P_j(\s'))=\d^i_j\d(\s,\s'), \\
&(P_i(\s),P_j(\s'))=0,\\
&(X^i(\s),X^j(\s'))=0.
\eea
%c1
See \cite{dirac} for a review of the Dirac procedure of constrained 
quantization.
%c1
Since 
\be
\del_\s X^i + \del_\tau X^j \cF_j{}^i = 
2 \pi \a' P^j \cF_j{}^i +  \del_\s X^j M_j{}^i,
\ee
the boundary condition is a constraint in the phase space
\be
\Phi^i( 0) = \Phi^i(\pi) =0, \quad i =0,1,\cdots, p,
\ee
where we have introduced the notation
\be
\Phi^i (\s) = 
2 \pi \a' P^j \cF_j{}^i +  \del_\s X^j M_j{}^i. 
\ee
Using the Hamiltonian 
\be
H_B=\frac{1}{4\pi\a'}\int d\s
\left( (\del_{\t}X)^2+(\del_{\s}X)^2\right),
\ee
it is easy to show that one requires also the constraints at $\s =0, \pi$
\footnote{
% p1
We thank
A. Bilal for a helpful discussion 
about the issue of equations of motion in determining 
the secondary constraints.
}
\be
\del^{2\n}_\s \Phi^i(\s)=0, \quad\quad \del^{2\n+1}_\s P^i (\s) =0, 
\quad\quad  \n=0,1, \cdots ,
\ee
and that these are all the second class constraints. We will denote
them by $\phi^{(\a k \n)}$, $\a =1,2$; $k=0,1,\cdots,p $;
$\n =0,1, \cdots $, 
\be \label{allcons}
\phi^{(1k \n)} = \del^{2\n}_\s \Phi^k, \quad 
\phi^{(2k \n)} = \del^{2\n+1}_\s P^k.
\ee
%c3 
These constraints are consistent with the 
explicit Fourier mode expansion of the fields $X^i$ and $P^i$.

One can then compute the Poisson matrix $C^{(\a k \n) (\b l \m)}$ 
of the constraints. The basic ones are the 
$C^{(\a k  0) (\b l 0)} \; $ 's :
\bea
(\Phi^k(\s), \Phi^l(\s')) &=& 2\pi\a' (\cF M)^{kl} 
[\del_\s \d(\s,\s') + \del_{\s'} \d(\s',\s) ],\\
(\Phi^k(\s), \del_{\s'} P^l(\s')) &=& 
M^{kl} \del_\s \del_{\s'} \d(\s,\s'), \\
( \del_{\s} P^k(\s), \del_{\s'} P^l(\s')) &=& 0
\eea 
and in general 
\be
% p2
C^{(\a k \n) (\b l \m)}(\s,\s')
=(\phi^{(\a k\n)}(\s), \phi^{(\b l\m)}(\s'))
% p2
=\del_\s^{2\n} \del_{\s'}^{2\m} C^{(\a k 0) (\b l 0)}(\s,\s').
\ee

A distinct feature of our case is that 
the constraints \eq{allcons} are imposed only at the boundary of
the open string.  As a result, the 
% p2
%c1
%usual form of the
%c1
Dirac bracket
%is slightly modified
%and is
should be given by
% p2
\bea \label{DB}
(A(\s), B(\s'))^* &=&(A(\s), B(\s')) \nn \\
&-& 
\sum_{\s^{''} \s^{'''}} (A(\s), \phi^{(\a k \n)} (\s^{''}))
C_{(\a k \n) (\b l \m)}(\s^{''}, \s^{'''} ) 
(\phi^{(\b l \m)} (\s^{'''}), B(\s')),  
\eea
where in place of an integral, we have a sum  $\s'', \s'''$
over the endpoints $0, \pi$. We also adopt the Einstein summation
convention for the indices $k,l, \n,\m$ unless otherwise stated.
With an obvious 
%c3 choice of 
labelling of the columns and rows of the matrix, 
the inverse $C_{(\a k  \n) (\b l \m)}(\s^{''}, \s^{'''}) $
is given by
\be \label{Cinv}
C_{(\a k \n) (\b l \m)}(\s^{''}, \s^{'''}) =
\pmatrix{0  & -(M^{-1})_{kl} R_{\n\m}(\s^{''}, \s^{'''}) 
\cr (M^{-1})_{kl} R_{\n\m}(\s^{''}, \s^{'''}) &  
2\pi\a' (\cF M^{-1})_{kl} S_{\n\m}(\s^{''}, \s^{'''})  }
\ee
with $R,S$ satisfying
\be
\sum_{\s^{''}} \del_\s^{2\n+1} \del_{\s''}^{2\m+1} \d(\s,\s'') 
R_{\m\k}(\s'' \s''') = \d^\n_\k \d_{\s \s'''}, \label{R} 
\ee
\be
\sum_{\s^{''}} \del_\s^{2\n} \del_{\s''}^{2\m}
[ \del_\s \d(\s,\s'') +\del_{\s''}\d (\s'',\s)]  
R_{\m\k}(\s'', \s''') =  \sum_{\s''}  
\del_\s^{2\n+1} \del_{\s^{''}}^{2\m+1} \d(\s,\s'') S_{\m\k}(\s'',\s''').
\label{S}
\ee
It follows immediately from the triangular form 
of $C_{(\a k\n) (\b l\m)}$ that 
\be
(P^i(\s), P^j(\s'))^* =0
\ee
is not modified.
To proceed with the computation of the rest of the Dirac brackets
among $X^i$ and $P^i$, one may try to invert the relations \eq{R}, 
\eq{S} to
determine the form of $R$ and $S$ and then use them in \eq{DB}.
However the explicit form of $R,S$ obtained this way are highly
singular and generally contain ambiguities.  Fortunately, there is a
trick to bypass these steps.  We will now show that  it is in fact
possible to compute the desired brackets using only the defining
relations \eq{R} and \eq{S}. The detailed form of $R, S$ is not needed. 

We first compute $(X^i(\s), X^j(\s'))^*$, the definition \eq{DB} gives
\bea \label{XX2}
(X^i(\s), X^j(\s'))^*  = -2 \pi \a' (M^{-1}\cF)^{ij} 
 \sum_{\s^{''} \s^{'''}}  
[\del_{\s''}^{2\n} \d(\s,\s'') R_{\n\m} (\s'',\s''') 
\del_{\s'''}^{2\m+1} \d(\s',\s''')   \nn \\
- \del_{\s''}^{2\n+1 } \d(\s,\s'') S_{\n\m}(\s'',\s''') 
\del_{\s'''}^{2\m+1}\d(\s',\s''') 
+ \del_{\s''}^{2\n+1 }  \d(\s,\s'')R_{\n\m}(\s'',\s''')
\del_{\s'''}^{2\m}\d(\s',\s''')
]. 
\eea
It is easy to see that  $(X^i(\s), X^j(\s'))^* =0$ for $\s, \s'$ not
both at the endpoints. Now we multiply  \eq{S} for the case $\n=0$ by 
$\sum_{\k}\sum_{\s'''} \del_{\s'''}^{2\k+1} \d(\s',\s''')$ 
and integrate over $\s$. We obtain
\bea
&&\sum_{\s^{''} \s^{'''}} 
\del_{\s''}^{2\m}\d(\s,\s'') R_{\m\k}(\s'',\s''') 
\del_{\s'''}^{2\k+1} \d(\s',\s''') |^{\s=\pi}_{\s=0} = \nn\\
&& \sum_{\s^{''} \s^{'''}}  \del_{\s''}^{2\m+1} 
\d(\s,\s'') S_{\m\k}(\s'',\s''') 
\del_{\s'''}^{2\k+1}\d(\s',\s''') |^{\s=\pi}_{\s=0},
\eea
where we have used
$\int d\s  \del_{\s''}\d(\s'',\s) =  \del_{\s''} \int d\s \d(\s'',\s)
=0$ to get rid of the second term on the left hand side of \eq{S}.
Therefore the first and second term in \eq{XX2} cancel each other in 
$(X^i(\pi)- X^i(0), X^j(\s'))^*$ and we are left with
\bea 
(X^i(\pi)- X^i(0), X^j(\s'))^* &=&  -2 \pi \a' (M^{-1}\cF)^{ij} 
\sum_{\s^{''} \s^{'''}}  \del_{\s''}^{2\n+1} \d(\s,\s'')
R_{\n\m}(\s'',\s''')\del_{\s'''}^{2\m}\d(\s',\s''') 
|^{\s=\pi}_{\s=0} \nn\\ 
&=& -2 \pi \a' (M^{-1}\cF)^{ij} \cdot 
(\d_{\s' 0} +\d_{\s' \pi} ), 
\label{pi-0}
\eea
where we have used in the last step above 
%c1 
\be \label{amb0}
\sum_{\s'' \s'''} \del_{\s''}^{2\n+1} \d(\s,\s'')
R_{\n\m}(\s'',\s''')\del_{\s'''}^{2\m}\d(\s',\s''') 
|^{\s=\pi}_{\s=0} 
= \d_{\s' 0} +\d_{\s' \pi} .
\ee
This can be obtained by taking the $\n=0$ 
case of \eq{R},
multiplying it with 
$\sum_{\k}\sum_{\s'''} \del_{\s'''}^{2\k} \d(\s',\s''')$ 
and integrating over $\s$; and
%c1 
note that
\be \label{amb}
\int_\s \sum_{\s'''} \d_{\s \s'''} \d(\s',\s''') = 
\int_\s \sum_{\s'''} \d_{\s \s'''} \d(\s',\s) = 
\sum_{\s'''} \d_{\s' \s'''} = \d_{\s' 0} +\d_{\s' \pi}. 
\ee
Formally, if one exchange the order of integration and the sum in the
above, one gets zero times a delta function.
Therefore \eq{amb} calls for a better justification.
One way to justify it is to use a lattice 
regularization by replacing 
the interval $[0,\pi]$ by a lattice of $M$ equidistant
points with spacing $\e =\pi/M$, 
\be
\d(\s,\s') \rightarrow \frac{1}{\e}\d_{\s \s'} , \quad 
\int d\s  \rightarrow \e \sum_\s .
\ee 

Since we know \cite{CH} from the D-brane field theory considerations 
that 
\be
(X^i(0), X^j(0))^* = - (X^i(\pi), X^j(\pi))^*,  
\ee
it follows from \eq{pi-0} that the nontrivial Dirac bracket must be
\be \label{DBXX}
(X^i(0), X^j(0))^* = - (X^i(\pi), X^j(\pi))^* = 
-2 \pi \a' (M^{-1}\cF)^{ij} .
\ee

Next, we compute $(X^i(\s), P^j(\s'))^*$. It is straightforward to
obtain
\be
(X^i(\s), P^j(\s'))^* = \eta^{ij} (\d(\s,\s') - Q(\s,\s')),
\ee
where
\be
 Q(\s,\s') = \sum_{\s'' \s'''} \del_{\s''}^{2\m+1} 
\d(\s, \s'') R_{\m\k}(\s'', \s''')
\del_{\s'''}^{2\k+1} \d(\s''',\s').
\ee
Using the definition \eq{R} with $\n=0$,
it is now easy to show that 
\be
\del_\s Q(\s,\s') = \d_{\s 0} \cdot [\del_\s \d(\s,\s') ]_{\s=0} + 
\d_{\s \pi} \cdot [\del_\s \d(\s,\s')]_{\s =\pi} 
\ee
and hence $\dt(\s,\s') \equiv \d(\s,\s') - Q(\s,\s')$ has the desired
property of having a vanishing derivative when one of its arguments
is at the boundary;
% p1
and $\dt(\s,\s')$ is just $\d(\s,\s')$ when
both arguments are away from the boundary.
The Fourier expansion of $\dt(\s,\s')$ can be found in \cite{CH}. 

%c2
Needless to say, some of the steps presented above are a little
formal. The consistent quantization using modes \cite{CH} provides a 
more concrete basis of our calculation here 
and can be viewed as  a complementary viewpoint of the same
quantization. From a practical point of view, the modes commutation
relations are more useful to perturbative string calculations.  
%c2

\section{Inclusion of Fermions}

In a generic  background, the fermionic part of a
% p1
RNS
open string 
\footnote{ We use the convention of 2-dimensional 
spinor algebra in \cite{LT}. In particular, 
$\psi = \pmatrix{\psi_+ \cr \psi_-}$, $ \rho^0 =\pmatrix{0&1 \cr -1 &0}$,
$\rho^1 =\pmatrix{0&1 \cr 1 &0}$.
}
gets additional couplings, for example 
%c3 \cite{open} 
\be
\frac{-i}{4\pi \a'} \int B_{ij} \psib^i \e^{\a\b}\rho_\a \del_\b \psi^j.
\ee
The situation is much
simpler for our case with the flat background \eq{flat} 
and the complete supersymmetric Lagrangian can be easily 
written down. 
The equation of motion is not modified
\be
\delb \psi_+^i = \del \psi_-^i =0,
\ee
where  $\del = \del_\t+ \del_\s, \delb = \del_\t- \del_\s$.
The open string worldsheet in the type \II and type 0 string theories
has the supersymmetry
\be \label{susy}
\d X^{\mu}=\bar{\epsilon}\psi^{\mu}, \quad
\d\psi^{\mu}=-i\r^{\a}\del_{\a}X^{\mu}\epsilon
\ee
with the preserved supersymmetry parametrized by
\be \label{goodSUSY}
\e_+ = \l \e_-, \quad \l =\pm 1.
\ee
The supersymmetric boundary condition compatible with 
\eq{BC1} is 
\footnote{
We thank V. Schomerus for
% p2
a
% p2
useful email exchange about this boundary
condition. A different boundary condition was used in \cite{CH} which
corresponds to a nonsupersymmetric D-brane.
} 
\bea
&\psi_+^i (\eta + \cF)_i{}^j + \l  \psi_-^i (\eta - \cF)_i{}^j =0, \quad
i,j =0,1,\cdots, p, \label{fBC1}\\
&\psi_+^a +\l   \psi_-^a =0, \quad a=p+1,\cdots, 9, 
\label{fBC2}
\eea
for  $\s=0, \pi$.

%c3 There is no difficulty to carry out 
The constrained quantization on the fermions can be similarly performed
%c3 the constrained quantization 
as in the bosonic case above. 
However, since we know that $\psi^i$ and $X^i$ are
related by supersymmetry and we have already derived the commutation
relations for the $X^i$'s, it is perhaps more interesting not to repeat
the Dirac quantization here, but instead to utilize the power of 
worldsheet supersymmetry to derive the Dirac bracket for
the $\psi^i$'s directly.  This approach is expected to be particularly
useful in more complicated cases. For example in the case \cite{CHK} 
where the
background is not constant but weak enough so that one can still 
determine the leading nontrivial commutation relation for the open
string; and in the case of supersymmetric $AdS_k\times S^k$ 
background \cite{work} relevant for the AdS/CFT correspondance. 

% p1
%Given that $(X^\mu(\s),X^\nu(\s'))^* = D^{\mu \nu}(\s,\s')$ is a field
%independent matrix, we can apply supersymmetry and obtain
%\be
%(\psi^\mu(\s), X^\nu(\s'))^* =  (\psi^\nu(\s), X^\mu(\s'))^*,
%\ee
%i.e. $(\psi^\mu(\s), X^\nu(\s'))^*$ is symmetric in $\mu\nu$. 
%It is possible to show that $(\psi^\mu(\s), X^\nu(\s'))^*$ is in fact
%again a field independent matrix and hence on applying \eq{susy}
%c1 
% Because there is no interaction term between $X^{\mu}$ and $\psi^{\mu}$
%in the action when $\cF$ is constant, 
It is easy to see that $(\psi^\mu(\s), X^\nu(\s'))^*=0$. 
The reason is that the original Poisson bracket 
$(\psi^\mu(\s), X^\nu(\s'))$ is zero and the constraints never mix the
bosonic and fermionic fields, 
so this Poisson bracket is not modified by the Dirac procedure. 
%c1
Applying \eq{susy}, we obtain
\be
0= \d (\psi^\mu, X^\nu)^* = \pmatrix{
[\;  i ( \del X^\mu, X^\nu)^* +  (\psi_+^\mu,\psi_+^\nu)^*  
 -\l (\psi_+^\mu,\psi_-^\nu)^* ] \; \cdot \e_-\cr 
[\;  -i ( \delb X^\mu, X^\nu)^* -  (\psi_-^\mu,\psi_-^\nu)^* 
 +\l (\psi_-^\mu,\psi_+^\nu)^* ] \; \cdot \l \e_-}.
\ee
Now we will concentrate on the modified boundary condition
\eq{fBC1}. The Dirac bracket relations for the tranverse
directions can be obtained simply  by setting $\cF=0$ in the following. 
Substituting \eq{fBC1} into the above, we obtain
\be \label{psipsi}
(\psi_+^i,\psi_+^j)^* = \frac{-i}{2} ( \del X^i, X^k)^* (\eta-\cF)_k{}^j, 
\quad\quad
(\psi_-^i,\psi_-^j)^* =\frac{-i}{2} ( \delb X^i, X^k)^* (\eta+\cF )_k{}^j.
\ee
At $\s=0, \pi$, the use of \eq{dXwithP} allows one 
to express $\del_\s X^i$ and  $\del_\t X^i$ in
terms of $P^k$
\be
\del_\s X^i = -2\pi \a' P^k (\cF M^{-1})_k{}^i, \quad\quad
\del_\t X^i = 2\pi \a' P^k (M^{-1})_k{}^i .
\ee
Thus
\bea
& (\del X^i (\s), X^j(\s'))^* = 2\pi  \a' (M^{-1}(\eta+\cF))^{ij}
\dt(\s,\s'), \\
&  (\delb X^i (\s), X^j(\s'))^* = 2\pi  \a' (M^{-1}(\eta-\cF))^{ij}
\dt(\s,\s'),
\eea
and so (\ref{psipsi}) implies
\be \label{DBpsi}
(\psi_+^i(\s),\psi_+^j(\s'))^* =  
(\psi_-^i(\s),\psi_-^j(\s'))^* =-i\pi \a'  \eta^{ij}\dt(\s,\s'),
\ee
with $\s$ 
%c1 or %c1
and $\s'$ at the boundary.
It is also easy to see that in the interior of the string, the Dirac
bracket is not modified and takes the same form as
\eq{DBpsi}. Therefore the fermion commutator is not modified 
% p1
(except for the modification of the delta function)
by the presence of $\cF$ and takes the standard form.

\section{Remarks}

% p1
In the quantization procedure of Dirac,
Eq.(\ref{amb}) is the only ambiguity.
But as mentioned in Sec.\ref{necess}, it is inconsistent
to say that (\ref{amb}) vanishes.
Also note that the ambiguity resides only in the commutation
relations for the endpoint coordinates, the quantization for
the spacetime coordinates in the interior of the opens string 
has no ambiguity, and their commutation relations
are the same as if $\cF=0$. 
%The fact that different regularizations
%may lead to different quantizations in this case was first
%pointed out by Seiberg \cite{Seiberg}.
The 
%c1
lattice
%c1
regularization we choose there gives the same result
we obtained in \cite{CH}. It leads to the conclusion that
a D-brane in the $B$ field background has a noncommutative
worldvolume in the sense that its worldvolume theory is
the SYM theory living on a noncommutative space.

Upon quantization, the Dirac bracket becomes the (anti)commutator of the
operators. From the point of view of the open string, 
the modifications to the commutation relations is ``not very
much''. But as discussed in \cite{CH}, since the endpoint of the open
string is living on the D-brane, the D-brane worldvolume becomes
a noncommutative one.
%c1
It is well known that 
a noncommutative manifold can be described
% p1
in the dual language of
the algebra of functions living on it, one
immediately arrives at the conclusion that the D-brane worldvolume
field theory is a noncommutative one, i.e., one with a modified
multipication. In fact it is easy to check that our modified 
% p2
commutators \eq{DBXX}
% and \eq{DBpsi}
% p2
are equivalent to 
the following Moyal product \cite{moyal}
% p1: I can't find the ref.
%,kontesvich} 
defined on the D-brane worldvolume \cite{CDS},
\be \label{moyal}
f(\xi) \star g(\xi') = 
\exp(\Theta^{ij}\frac{\del}{\del \xi^i}\frac{\del}{\del \xi^{'j}}) 
f(\xi) g(\xi') |_{\xi=\xi'},
\ee
where $\xi^i$, $i= 0,\cdots, p$ are 
the D-brane worldvolume coordinates and 
\be
\Theta^{ij} = \pm i \pi \a' (\cF M^{-1})^{ij}.
\ee

The 
%c1
noncommutative
D$p$-brane worldvolume theory
% p1
%c1 is given by the
takes functionally the same form as the 
SYM$_{9+1}$ theory dimensionally reduced
to $p+1$ dimensions, with the usual product of fields replaced by  
the Moyal product \eq{moyal}, and with the spacetime integral replaced
by  a cyclic trace.
%\be
%L = ...
%\ee
%c1 
It has been argued \cite{Seiberg}
% p2
%,Witten}
% p2
that
in the DLCQ limit of $\eps\rightarrow 0$ with
\bea
&X^{\mu}\sim\eps, & B\sim 1/\eps^2, \\
&\a'\sim\eps, & g_s\sim\eps^{3/2},
\eea
where $g_s$ is the string coupling constant,
the SYM Lagrangian is exact and hence deserves attention.

When this work is nearly finished, we learned that
another group \cite{AAS1} has also employed Dirac's procedure
for the same problem, but their final result
is different from ours. The reason is presumably that
they chose a different regularization.

\newpage
\section*{Acknowledgment}

We thank A. Bilal, M. Cederwall, Y.-C. Kao, D. Matalliotakis, 
R. Russo and V. Schomerus for helpful
discussions. The work of C.S.C. is supported by the Swiss National
Science Foundation. The work of P.M.H. is supported in part by the
National Science Council, Taiwan, R.O.C.

%\newpage

\ed